# Solubility limit of Ge Dopants in AlGaN: a Chemical and Microstructural Investigation down to the Nanoscale

C. Bougerol,[†] E. Robin,[‡] E. Di Russo,[§] E. Bellet-Amalric,[‡] V. Grenier,[‡] A. Ajay,[‡,⊥] L. Rigutti,[§] and E. Monroy[‡,*]

[†] Univ. Grenoble-Alpes, Institut Néel-CNRS, 25 av. des Martyrs, 38000 Grenoble, France
[‡] Univ. Grenoble-Alpes, CEA, IRIG, 17 av. des Martyrs, 38000 Grenoble, France
[§] UNIROUEN, CNRS, Groupe de Physique des Matériaux, Normandie Université, 76000 Rouen, France

* Corresponding author: eva.monroy@cea.fr

[⊥] Current affiliation: Walter Schottky Institute and Physics Department, Technical University of Munich, 85748 Garching, Germany

OrcID:
Catherine Bougerol: 0000-0002-4823-0919
Eric Robin: 0000-0002-5596-2640
Enrico Di Russo: 0000-0003-3829-6567
Edith Bellet-Amarlic: 0000-0003-2977-1725
Vincent Grenier: 0000-0003-2781-0559
Akhil Ajay: 0000-0001-5738-5093
Lorenzo Rigutti: 0000-0001-9141-7706
Eva Monroy: 0000-0001-5481-3267

**ABSTRACT:** Attaining low resistivity $Al_xGa_{1−x}N$ layers is the keystone to improve the efficiency of light emitting devices in the ultraviolet spectral range. Here, we present a microstructural analysis of $Al_xGa_{1−x}N$:Ge samples with $0 \leq x \leq 1$, and nominal doping level in the range of $10^{20}$ cm$^{−3}$, together with the measurement of Ge concentration and its spatial distribution down to the nm scale. $Al_xGa_{1−x}N$:Ge samples with $x \leq 0.2$ do not present any sign of inhomogeneity. However, samples with $x > 0.4$ display µm-size Ge crystallites at the surface. Ge segregation is not restricted to the surface: Ge-rich regions with a size of tens of nanometers are observed inside the $Al_xGa_{1−x}N$:Ge layers, generally associated with Ga-rich regions around structural defects. With this local exceptions, the $Al_xGa_{1−x}N$:Ge matrix present an homogenous Ge composition which can be significantly lower than the nominal doping level. Precise measurements of Ge in the matrix provide a view of the solubility diagram of Ge in $Al_xGa_{1-x}N$ as a function of the Al mole fraction. The solubility of Ge in AlN is extremely low. Between AlN and GaN, the solubility increases linearly with the Ga mole fraction in the ternary alloy, which suggests that the Ge incorporation takes place by substitution of Ga atoms only. The maximum percentage of Ga sites occupied by Ge saturates around 1%. The solubility issues and Ge segregation phenomena at different length scales likely play a role in the efficiency of Ge as n-type AlGaN dopant, even at Al concentrations where Ge DX centers are not expected to manifest. Therefore, this information can have direct impact in the performance of Ge-doped AlGaN light emitting diodes, particularly in the spectral range for disinfection ($\approx$ 260 nm), which requires heavily-doped alloys with high Al mole fraction.

**KEYWORDS:** *AlGaN, germanium, dopant, solubility*

# ■ INTRODUCTION

Following the generalization of blue and white light emitting diodes (LEDs) based on InGaN/GaN semiconductors, major research efforts are now oriented towards the development of LEDs operating in the ultraviolet (UV) region.[1] UV lamps present extensive applications not only in germicidal systems and wastewater purifiers,[2,3] but also as curing lamps[4] and in the domain of phototherapy.[5] Demand for disinfection of air, surfaces or medical instruments has boosted during the 2020 pandemics[6], and it is expected to increase further in light of the growing need for safe drinking water and food sterilization. Mercury lamps are the most widely used devices for such applications, but there is a strong motivation to replace them by LEDs,[7] which are more eco-friendly and compact, can switch faster and potentially present longer lifetime. AlGaN is the most promising and mature technology for semiconducting UV emitters, but the LED efficiency results are quite modest.[1] The performance is mainly limited by severe material issues, such as the high resistivity of the contact layers due to the low activation rate of dopants in high aluminum content alloys.

Silicon has long been the preferred and most investigated n-type dopant in GaN. However, at high doping level, Si induces tensile stress responsible for film cracking and surface roughening.[8–10] In the case of $Al_xGa_{1-x}N$ alloys, it is difficult to achieve highly conductive Si-doped layers with x > 0.7,[11–13] which is only partially explained by the higher activation energy. High resistivity has also been attributed to carrier compensation by deep level defects, including deep Si DX centers, which are formed when the shallow donor impurity undergoes a large bond-rupturing displacement and becomes a deep acceptor. Different calculations support that Si forms a deep DX center in $Al_xGa_{1-x}N$,[14–17] with an expected onset of DX behavior occurring somewhere in the range of 0.24 < x < 0.94. Experimental results also suggest that Si behaves as a DX center in $Al_xGa_{1-x}N$, with onset in the range of 0.42 < x < 0.84.[18–20]

In this context, germanium (Ge) having an atomic radius similar to Ga should induce less stress, and appears therefore as an alternative n-type dopant, especially for applications requiring concentrations above the Mott transition ($\approx 10^{19}$ cm$^{-3}$), e.g. plasmonics,[21] intersubband devices,[22,23] or the implementation of the tunnel junctions that are currently used to reduce the resistance of the contact to the p-type region in GaN-based LEDs.[24,25] Indeed, crack free GaN:Ge films with dopant concentration above $10^{19}$ cm$^{-3}$ were successfully grown by molecular beam epitaxy (MBE)[26,27] or metalorganic vapor phase epitaxy (MOVPE)[10,21] and showed a smooth surface morphology. Regarding the electronic properties, Ge is a shallow donor in GaN with an ionization energy of about 30 meV.[13,27,28] Comparing the length of the Ge-N bond ($\approx$ 1.89-2.00 Å)[29,30] with that of the Ga-N and Al-N bonds (1.95 Å and 1.89 Å, respectively), Ge should occupy the Ga or Al lattice sites in $Al_xGa_{1-x}N$ causing little lattice distortion. Ajay, et al.[27] reported that the carrier concentration in GaN:Ge scales linearly with the Ge concentration. However, when substituting Al for Ga, the ionization energy of germanium increases and the carrier density at constant Ge concentration decreases gradually.[31] This has been theoretically predicted to occur for Al mole fractions higher than 50% due to a DX transition involving the formation of a deep acceptor state (DX center) and subsequent self-compensation.[15] However, very recently, Bagheri, et al. carried out Hall measurements which did not observe any of acceptor states.[32] They proposed that the DX center would correspond to a deep donor transition whose ionization energy increases with the Al content up to 160 meV for $Al_{0.6}Ga_{0.4}N$, and concluded that high conductivity in Ge-doped Al-rich $Al_xGa_{1-x}N$ is in principle achievable, even for pure AlN.

At this point, the issue of Ge solubility in $Al_xGa_{1-x}N$ and the effective dopant incorporation in the entire compositional range between GaN and AlN appears very relevant. The formation of precipitates in the case of heavy doping has been reported in several semiconductors families, both in bulk materials[33,34] and thin films[35]. In the case of GaN:Ge grown by MBE, Hageman, et al.[26] observed the presence of secondary phases, namely $Ge_3N_4$, and the presence of pure Ge was identified by X-ray diffraction (XRD) for dopant concentration above $4\times10^{20}$ cm$^{-3}$. More recently, in $Al_xGa_{1-x}N$:Ge samples with x < 0.15 and Ge concentrations > $10^{20}$ cm$^{-3}$, we observed by particle-induced X-ray emission (PIXE) the formation of crystallites which were identified as metallic Ge by scanning electron microscopy (SEM) coupled to energy dispersive X-ray analysis (EDX).[31] Besides the presence of these secondary phases, diffusion of dopants along structural defects may occur, which would have an impact on the concentration of Ge atoms available for doping. Segregation of dopants along threading dislocations has been recently reported in Mg-doped GaN[36] and III-As[37], or at anti-phase domain boundaries in cubic GaN:Ge.[38] Based on the above considerations, the potential of Ge as a dopant in $Al_xGa_{1-x}N$ can only be validated by studies that combine an accurate determination of the Ge concentration and a careful analysis of its spatial distribution down to the nanometer scale. In order to address this issue, we have grown a series of $Al_xGa_{1-x}N$ ($0 \leq x \leq 1$) samples with different Ge concentrations. We report here the microstructural analysis of the different samples based on a combination of XRD and quantitative EDX coupled to high-resolution scanning transmission electron microscopy (HR-STEM) and atom probe tomography (APT). From these studies, we extract the solubility limit of germanium in $Al_xGa_{1-x}N$ ($0 \leq x \leq 1$).

## ■ RESULTS AND DISCUSSION

$Al_xGa_{1-x}N$ ($0 \leq x \leq 1$) samples with a thickness of ≈ 675 nm and various Ge concentrations were grown by plasma-assisted MBE on AlN-on-sapphire templates, as described in the Methods. The Al mole fraction and Ge content were analyzed by EDX, with the results presented in Table 1. For confirmation, selected samples were measured by secondary ion mass spectrometry (SIMS, data included in the table). Deviations between both techniques can be due to the use of different standards. In this study, we will focus on EDX results, with error bars extracted from measurements at various points of the sample and with two accelerating voltages (5 kV and 15 kV). Samples grown with a Ge cell temperature $T_{Ge}$ = 1011°C present an average Ge concentration [Ge] = (2.2±0.2)×10$^{20}$ cm$^{-3}$, whereas the two samples grown with $T_{Ge}$ = 928°C show [Ge] = (2.85±0.55)×10$^{19}$ cm$^{-3}$, consistent with the expected exponential reduction of the Ge atom flux. However, as the resolution limit of EDX is about 10$^{19}$ cm$^{-3}$, we have decided to focus mainly on samples grown with $T_{Ge}$ = 1011°C, i.e. with [Ge] > 10$^{20}$ cm$^{-3}$. The information on AL10GE1 and AL20GE1 is included in the table to show that we are well above the resolution limit of the system.

In the heavily doped samples, the average Ge composition does not present any particular trend as a function of the Al mole fraction of the samples, which might suggest an efficient incorporation of Ge into $Al_xGa_{1-x}N$. However, things look very different after a careful structural analysis. In $Al_xGa_{1-x}N$:Ge samples with $x > 0.4$, top-view SEM observations reveal the presence of μm-size crystallites at the surface, whose density tends to increase with the Al content of the samples. In contrast, $Al_xGa_{1-x}N$:Ge samples with $x \leq 0.2$ present smooth surfaces, even when varying the growth temperature by ±30°C (+30°C for samples AL10GE2B and AL20GE2B, and −30°C for AL20GE2C). EDX measurements show that the crystallites consist of pure Ge, as illustrated in Figure 1a.

For AlN:Ge samples, before HCl treatment, the surface appears decorated with large (5-10 μm in diameter) drops consisting of metallic Al and Ge (see Figure 2a). Germanium appears systematically located at the periphery of the droplets (see zoomed view in Figure 2c). EDX spectra taken in an area between droplets, shown in Figure 2c-1, displays the K lines of N (at 0.39 keV) and Al (1.49 keV). In contrast, a similar analysis in the central area of the droplet, in Figure 2c-2, shows only the K line of Al, without contribution of nitrogen or oxygen, which confirms that droplets are pure Al, accumulated as a result of the metal-rich growth conditions. Finally, the red area at the periphery of the droplet presents only the L lines of Ge (1.04, 1.07, 1.18 and 1.22 keV), without any indication of nitrogen or oxygen (Figure 2c-3). We therefore conclude that these areas are metallic Ge. After HCl treatment, aluminum is dissolved and only metallic germanium remains (see Figure 2b). This clear separation of Al and Ge is consistent with their reported immiscibility.[39]

Interestingly, we did not observed $Ge_3N_4$ inclusions in any of the samples, in contradiction with data reported by Hageman, *et al.*[26] and Fireman, *et al.*[40]. In the case of $Al_xGa_{1-x}N$:Ge with x < 1, this might be explained by the fact that $Ge_3N_4$ is dissolved by the HCl after-growth cleaning.[41] However, sample AL100GE2 was not treated by HCl and the only Ge crystals present at the surface are pure metal, as shown in Figure 2. The absence of $Ge_3N_4$ could be due to the fact that $Ge_3N_4$ starts to evaporate at 600°C and decomposes above 800°C.[10] Therefore, a certain decomposition is expected at our growth temperature (720°C).

The presence of pure Ge at the surface in highly-doped $Al_xGa_{1-x}N$ (x ≥ 0.4) samples implies that the average Ge concentrations measured by EDX or SIMS might not give information about the real doping level. Measuring the Ge concentration in the matrix, away from the Ge crystallites, is mandatory and can be only achieved with nanometer scale resolution methods such as EDX. In Table 1, for samples showing Ge crystallites at the surface, the Ge concentration labelled "matrix" was measured in a region far from the crystallites. Note that the "matrix" concentration represents an upper limit. On the one hand, there might be Ge clusters in the matrix, which would contribute to the signal. On the other hand, the Ge crystallites at the surface are thick and backscattered electrons might excite them from a certain distance and create a secondary fluorescence. However, this contribution is probably negligible (we estimate a maximum contribution around 10% of the signal measured in the matrix), as shown by the weak signal of Ge in AlN.

To get more insight in the Ge distribution in the matrix, EDX chemical maps were also obtained for sample AL50GE2 prepared in cross section and observed by HR-STEM, with the result illustrated in Figure 3. The chemical map shown in Figure 3b, corresponding to the zone outlined with a rectangle in Figure 3a, shows a marked inhomogeneity in the Ge distribution associated with a non-uniform distribution of Al and Ga: nanometer-size Ge-rich regions are located on top of Ga-rich areas along the growth direction, both being preferentially surrounded by Al-rich regions. Phase separation is unlikely in $Al_xGa_{1-x}N$,[42] but $Al_xGa_{1-x}N$ demixing associated to heavy doping was reported for Si-doped $Al_xGa_{1-x}N$ grown by plasma-assisted MBE by Somogyi, et al.[43] In this case, silicon clusters appear associated with domains with higher Al mole fraction. It was argued that the phenomena might be associated to a modification of the growth process linked to a surfactant effect of Si.[43] However, under the Ga-rich conditions required for the self-regulated growth of planar GaN, the presence of an additional Si or Ge flux does not have any significant effect on the growth kinetics.[27,44] In the case of Ge, strain is not expected to play a major role, since the bond lengths of Ge-N, Al-N and Ga-N are relatively similar. However, the local organization of Al, Ga and Ge-rich regions might be related to the different bonding energies: the Al-N bonding energy (2.88 eV) is much stronger than that of Ge-N (2.66 eV) and Ga-N (2.20 eV),[45] which might favor that Ge replaces Ga in the AlGaN lattice. Also, in view of the low solubility of Ge in the AlN matrix (see AL100GE2 in Table 1), Ge atoms might present a tendency to minimize the number of neighboring Al atoms. It is difficult to precisely evaluate the impact of these Ge-rich clusters on the Ge concentration really available for doping the AlGaN matrix. However, as their volume is very small compared to the Ge crystallites at the surface, their impact will likely be minor.

To explore the arrangement of Ge crystallites or Ge inclusions on or in the AlGaN matrix, out-of-plane X-ray diffraction (XRD) measurements where performed for all the samples. As illustrated in Figure 4a, samples with and Al content below 20% display only diffraction peaks assigned to the AlN-on-sapphire substrate and the $Al_xGa_{1-x}N$ layer. On the contrary, for the samples with high Al content, additional peaks appear in the θ−2θ diffractogram (see AL60GE2 in Figure 4a). Some can be unambiguously attributed to crystalline Ge [Ge(111) at 2θ = 27.3° and Ge(220) at $2\theta = 45.30°$]. These Ge-related peaks are too narrow ($\approx 0.2°$) to stem from nanometer-scale crystals (e.g. the inclusions observed in cross-section STEM images like in Figure 3), i.e. they can only originate from the Ge crystallites observed at the surface (Figure 1). They exhibit a preferential orientation (higher intensity) along the [220] direction although their mosaic spread ($\Delta\omega \approx 1°$) is much higher than that of the $Al_xGa_{1-x}N$-related lines ($\Delta\omega \approx 0.06$ degrees). The other minor diffraction peaks could not be assigned (they could not be explained by the presence of $Ge_3N_4$).

In order to investigate the in-plane epitaxial relation of the Ge crystallites we performed in-plane $2\theta_\chi$–$\phi$ scans along the main AlN <11-20> and <1-100> directions for sample AL60GE2, as displayed in Figure 4b. In addition to the diffraction peaks from the substrate and $Al_xGa_{1-x}N$, we observe all the Ge reflections with different intensity, depending on the scan direction. This trend points to an in-plane preferential direction, which that can be determined by performing $\phi$ scans for different reflections. In Figure 4c, we superimposed the scans for the <11-20> direction of AlGaN (aligned with the <3-300> direction of the sapphire substrates) and the <220> direction of Ge (most intense peak). The Ge crystallites follow the 6 fold symmetry of nitrides with very broad peaks at 0° and 60°. However, between these two directions, the intensity is higher than the background level. Therefore, we conclude that some of the Ge crystallites present a preferential epitaxial orientation, as shown by the presence of clear facets (see top inset of Figure 1), whereas others are randomly oriented.

The above-described XRD analysis could not be performed in the case of AL100GE2. In this case, the Al excess of the as-grown sample leads to the observation of intense X-ray reflections associated with face-centered cubic aluminum (see $\theta$–$2\theta$ scan in Figure S1 in the Supporting

Information), which mask the presence of germanium. After chemical cleaning to remove aluminum, the only visible reflections were those of AlN and sapphire.

To further investigate the possibility of nanometer-scale Ge clustering or Ge inhomogeneities in the AlGaN matrix, we have performed atom probe tomography studies of samples AL20GE2C and AL50GE2. This technique give us access to the distribution of Ge at a scale below the spatial resolution limit of EDX. Figure 5a shows the entire three-dimensional (3D) reconstruction of one of the APT specimens from AL50GE2 showing the position of Ga, Al and Ge ions. Figure 5b displays a mass/charge spectrum recorded during the APT measurement, where the peaks corresponding to Ge ions are highlighted, confirming that they are clearly resolved. We performed a statistical analysis on Ge 3D distribution over the whole specimen volume, comparing it to an artificial alloy with Ge, Ga and Al atoms mixed in a random manner. From the similarity of the results illustrated in Figure 5c, we conclude that there is no clustering at the nanometer scale, and the Ge atoms are randomly distributed. Note that the presence of clusters at the level of 2-5 atoms would be below the detection limit of APT.

These results confirm the uniform distribution of Ge in the $Al_xGa_{1-x}N$ matrix at the nanometer scale. In the three APT specimens extracted from AL50GE2, it was not possible to observe the Ge-rich inclusions identified by STEM/EDX, with a size about 20-30 nm in diameter and up to 100 nm in length. This is explained by the fact that such inclusions were scarce and randomly dispersed in the matrix, and the field of view of APT is much smaller than that of STEM/EDX (the APT tip size is ≈ 100 nm in diameter and ≈ 300 nm in length, to be compared with the ≈ 5-μm-long, ≈ 150-nm-thick STEM/EDX lamella). Nonetheless, some APT experiments show the presence of structural defects, probably threading dislocations, which appear as a linear inhomogeneity in the Ga and Al mole fractions (see arrows in Figure 6). Ge-rich regions appear connected to some of the Ga-rich areas. However, this association is not systematic, Ge remains homogeneous around some of these structural defects. A possible explanation is that Ge propagates only along certain types of threading dislocations. In this line, Zhang, *et al.*[46] has reported Ge regions associated to screw-type threading dislocations, which are a minority in the layers (most threading dislocations are edge type)[47].

The results in Figures 5 and 6 describe the Ge distribution in the matrix of an Al-rich sample, namely AL50GE2. For the sake of completeness, we include an APT analysis of AL20GE2C in Figure S2 of the Supporting Information. In this sample, with lower Al mole fraction, the incorporation of Ge is homogeneous, with the exception of a local enrichment at some threading dislocations, similarly to the observations in Figure 6.

In summary, Ge incorporates homogeneously in the $Al_xGa_{1-x}N$ matrix up to a certain limit, which depends on the Al content of the layers, and the excess of Ge segregates at the surface and forms crystallites of pure Ge. To assess the incorporation limit as a function of the Al mole fraction, Figure 7 presents a summary of the Ge concentration measured by EDX in all the samples. In the case of layers with Ge crystallites, we have included the average concentration (red triangles) and the concentration in the matrix (blue diamonds). The average Ge concentration in $Al_xGa_{1-x}N$ remains constant, slightly above $2\times10^{20}$ cm$^{-3}$, in all the Al compositional range. In contrast, the Ge concentration in the matrix decreases strongly for $x > 0.4$, to reach $2.4\times10^{19}$ cm$^{-3}$ for $x = 1$, indicating an extremely low solubility of Ge in AlN. In fact, in the Al-Ge-N system, the incorporation of Ge atoms in the AlN lattice was only reported for a metastable solid solution in thin films grown by magnetron sputtering,[48] which decomposes upon annealing with the formation of Ge crystallites. A linear fit of the Ge content in the matrix in the case of saturated samples (with Ge crystallites) give us a view of the saturation threshold of Ge in $Al_xGa_{1-x}N$ (dashed line in Figure 7). From our data, the incorporation of Ge in AlN is negligible and the saturation threshold increases linearly with the Ga mole fraction of the ternary alloy, which would suggest that the incorporation of Ge in AlGaN takes place by substitution of Ga atoms. With this assumption, the maximum percentage of Ga sites occupied by Ge would saturate around 1%. Our diagram is consistent with the solubility limit around $4\times10^{20}$ cm$^{-3}$ for Ge in GaN reported by Fireman, *et al.*[40] and Hageman, *et al.*[26]

## ■ CONCLUSIONS

In summary, we have analyzed the incorporation of Ge in $Al_xGa_{1-x}N$ layers ($0 \leq x \leq 1$) grown by plasma-assisted molecular beam epitaxy, targeting a doping level in the range of $10^{20}$ cm$^{-3}$. Samples with an Al atomic fraction $x > 0.4$ display Ge crystallites at the surface. Inside those layers, we have also identified Ge-rich inclusions with a size of tens of nanometers, generally associated with Ga-rich regions around structural defects. With these local exceptions, the $Al_xGa_{1-x}N$:Ge matrix presents homogenous Ge composition. The Ge content in the AlN matrix is extremely low, and it increases linearly with the Ga mole fraction in the $Al_xGa_{1-x}N$ matrix, which suggests that the Ge incorporation takes place by substitution of Ga atoms. The maximum percentage of Ga sites occupied by Ge saturates around 1%. These solubility issues and the Ge segregation phenomena should play a role in the efficiency of Ge as n-type dopant, even at Al concentrations where Ge DX centers are not expected to manifest. In view of these results, the extracted solubility limit of Ge in $Al_xGa_{1-x}N$ can have direct impact in the performance of $Al_xGa_{1-x}N$-based UV light emitting diodes. In principle, the Ge content should be kept below these limits to prevent a degradation of carrier transport due to scattering at structural defects. Co-doping or the use of surfactant species should be explored to attain Ge concentrations beyond the solubility limit.

# ■ METHODS

The samples under investigation are Ge-doped $Al_xGa_{1-x}N$ ($0 \leq x \leq 1$) epitaxial layers with a thickness of 675 nm, grown by plasma-assisted MBE on 1-µm-thick commercial AlN-on-sapphire templates. The nitrogen cell parameters where tuned so that the active nitrogen flux was $\Phi_N \approx 0.5$ monolayers per second (ML/s) for all the samples. The growth was performed under slightly metal-rich conditions. In the case of $Al_xGa_{1-x}N$ (x < 1), the metal-to-nitrogen flux ratio was $(\Phi_{Al} + \Phi_{Ga})/\Phi_N \approx 1.1$, where $\Phi_{Al}$, and $\Phi_{Ga}$ are the atomic fluxes of Al and Ga. The aluminum cell temperature was fixed so that $\Phi_{Al} = x\Phi_N$, where x was the desired Al mole fraction, and the metal excess was provided by the Ga flux. During the growth, the surface morphology was monitored in real time by reflection high-energy electron diffraction (RHEED), verifying that the RHEED intensity remained constant and the RHEED pattern was streaky. To prevent any degradation of the surface during the cooling down process, the Ge-doped $Al_xGa_{1-x}N$ samples were metalized with Ga at the growth temperature. Once extracted from the MBE system, the samples were chemically cleaned with high-purity HCl for 5 min, to remove the metal excess on the surface. In the case of AlN, the substrate temperature is too low for Al to desorb from the growing substrate. Therefore, the Al flux was fixed so that $\Phi_{Al}/\Phi_N \approx 1.05$, and the sample was grown without interruptions. This leads to an accumulation of Al on the surface, so that it is not necessary to metalize the sample during the cooling process, and the Al excess can be removed *ex situ* by HCl cleaning. This Ge-doped AlN sample was characterized before and after HCl cleaning. The list of samples under study and their relevant growth parameters are summarized in Table I.

The samples were analyzed by XRD using a Rigaku SmartLab diffractometer. Out-of-plane scans were performed using a 2-bounce Ge (220) monochromator and a long plate collimator of 0.228° for the secondary optics. For the in-plane measurements the diffractometer was equipped with 0.5° collimators in the primary and secondary optics, and with a nickel filter to suppress the K$_\beta$ contribution. In both out-of-plane and in-plane configurations the signal is averaged over several mm$^2$.

This macroscopic scale investigation was combined with the analysis of the microstructure of all samples using EDX in order to map the Ge distribution and determine the chemical composition at the nanometer scale. Top-view experiments were carried out in an Ultra55 Zeiss SEM equipped with a Flat Quad 5060F annular detector from Bruker. The major difficulty for quantifying low concentrations by EDX is the fact that a small signal has to be extracted from a relatively high background. To overcome this problem, we have developed a new method based on the use of specific windows that act as X-ray filters leading to an enhancement of the signal-to-noise ratio in the energy range corresponding to the analyzed dopants. We have developed an analytical procedure for removing the background based on the normalization of the spectrum to an undoped reference spectrum. Finally, the main problem to achieve accurate quantitative EDX analysis is the correction of matrix effects, mainly electron stopping power and backscattering, as well as X-ray absorption that may affect X-ray generation and emission depending on the accelerating voltage, the energy of the considered X-ray line and the composition and mass-thickness of the sample. In this study, these effects were corrected using a patented method,[49] implemented in a homemade code (IZAC) and calibrated on reference samples of known composition and thickness. The entire procedure was checked on Mg- and Si-doped GaN, as well as on p-doped Ge thin films. Results are consistent with SIMS analyses, even for concentrations below $10^{19}$ cm$^{-3}$.[50]

In order to check the diffusion of Ge along threading dislocations and also further investigate the potential secondary phases, sample AL50GE2 was also prepared in cross section in Zeiss Crossbeam550 scanning electron microscope/focused ion beam (SEM/FIB). To extract the region of interest, the as-grown sample was first coated with a 100-nm-thick Pt layer using a GATAN PECS system. After depositing *in situ* 500-nm-thick Pt layer with GIS, a standard lift-out procedure was performed with $Ga^+$ ions accelerated at 30 kV. The lamella thinning process was carried out at 15 kV, with a final clean-up step at 2 kV. The lamella was subsequently studied by EDX coupled to high resolution scanning transmission electron microscopy (HR-STEM) using a FEI Themis probe corrected microscope operated at 200kV and equipped with Super-X detectors.

The presence of Ge inhomogeneities was further studied by laser-assisted APT adopting the same lift-out procedure as for the lamella specimen, followed by annular milling performed at 15 kV, with a final 2 kV clean-up step. SEM images of the final filed emitters were collected to measure both sample cone-angle and tip dimensions, in order to perform APT 3D reconstructions.[51] The samples were analyzed in a laser-assisted wide-angle tomographic atom probe (LaWaTAP) system equipped with an ultraviolet (343 nm) femtosecond laser, with a repetition rate of 100 KHz. The detection system was a custom-designed multi-channel-plate/advanced delay line detector (MCP/aDLD) improved for the multi-hit detection,[52] with an overall detection efficiency $\eta \approx 0.6$. APT has proven to be a valuable tool for the analysis of impurity distribution, segregation and clustering. However, it presents some drawbacks preventing an accurate determination of impurity concentration or alloy mole fraction in compound semiconductors.[53] Such compositional biases were not corrected in the APT data presented in this manuscript.

## ◼ ASSOCIATED CONTENT

**Supporting Information**

X-ray diffraction measurements of AL100GE2 and atom probe tomography analysis of AL20GE2C.

## ◼ ACKNOWLEDGEMENTS

This work is supported by the French National Research Agency via the GaNEX program (ANR-11-LABX-0014) and the UVLASE project (ANR-18-CE24-0014), by the Auvergne-Rhône-Alpes Region via the PEAPLE project, and by the RIN IFROST project by the European Union with the European Regional Development Fund (ERDF) and by Region Normandy.

■ REFERENCES

**Table 1.** Description of samples under study: Ge cell temperature ($T_{Ge}$), substrate temperature ($T_S$), Al content obtained from EDX and SIMS, and Germanium concentration measured by EDX and SIMS. In the case of EDX, error bars are extracted from measurements at various points of the sample and with two accelerating voltages (5 kV and 15 kV). In the case of SIMS, measurements were performed at a single point, and the error bars represent the fluctuations of the composition in depth.

| Sample | | $T_{Ge}$ (°C) | $T_S$ (°C) | Al mole fraction | | [Ge] (×10$^{19}$ cm$^{-3}$) | |
|---|---|---|---|---|---|---|---|
| | | | | EDX | SIMS | EDX | SIMS |
| AL0GE2 | | 1011 | 720 | 0 | 0 | 19.7±0.8 | 26±9 |
| AL10GE1 | | 928 | 720 | 0.102±0.004 | 0.084±0.002 | 3.4±0.7 | 6.3±0.3 |
| AL10GE2 | | 1011 | 720 | 0.098±0.004 | 0.080±0.002 | 23.5±1.0 | 39±5 |
| AL10GE2B | | 1011 | 750 | 0.104±0.004 | | 21.3±2.7 | |
| AL20GE1 | | 928 | 720 | 0.203±0.007 | | 2.3±0.7 | |
| AL20GE2B | | 1011 | 750 | 0.201±0.007 | | 24.6±3.9 | |
| AL20GE2C | | 1011 | 690 | 0.194±0.007 | | 22.9±4.0 | |
| AL40GE2 | average | 1011 | 720 | 0.450±0.011 | 0.484±0.005 | 25.2±2.5 | 18±6 |
| | matrix | | | | | 23.3±2.3 | |
| AL50GE2 | average | 1011 | 720 | 0.566±0.011 | | 19.9±1.8 | |
| | matrix | | | | | 19.0±2.5 | |
| AL60GE2 | average | 1011 | 720 | 0.592±0.011 | | 22.0±2.0 | |
| | matrix | | | | | 12.9±1.7 | |
| AL100GE2 | average | 1011 | 720 | 1 | | 20.2±1.8 | |
| | matrix | | | | | 2.4±0.2 | |

# Figures

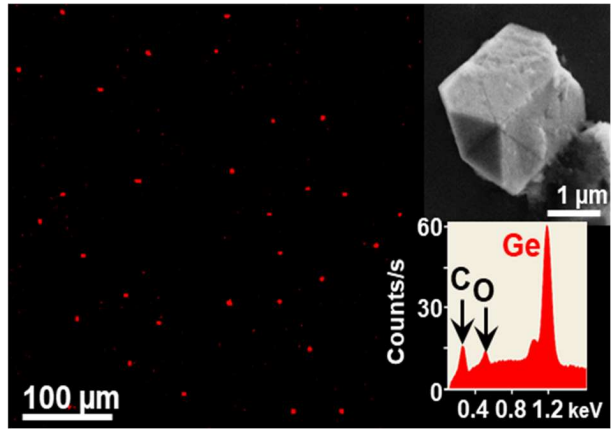

**Figure 1.** Top-view EDX map of sample AL50GE2 showing Ge crystallites in red. Top inset: Top-view SEM image of a single crystallite of about 2 µm in size (typical size: 1-3 µm). Bottom inset: EDX spectrum of a crystallite showing that it is pure Ge (C and O are surface contaminants).

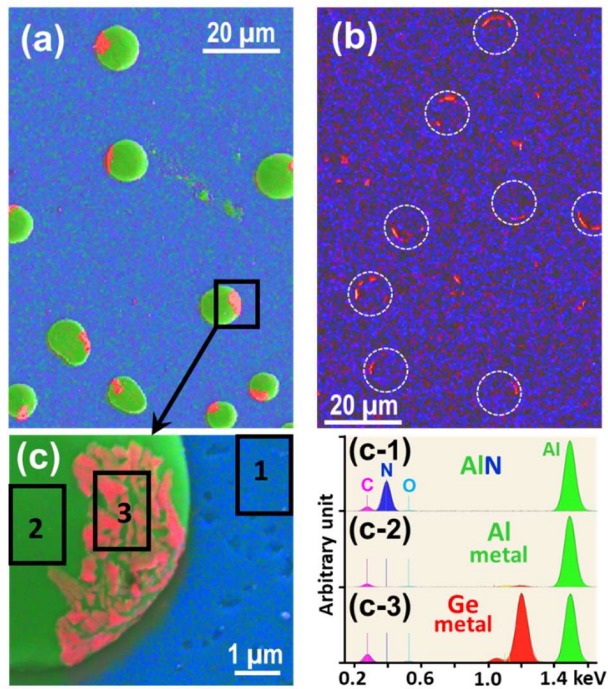

**Figure 2.** (a) Top-view EDX map of sample AL100GE2 before HCl treatment. (color code: Ge red, Al green, N blue). The surface is decorated with large droplets consisting of Al and Ge. (b) Top-view EDX map of sample AL100GE2 after HCl treatment. (color code: Ge red, N blue). The Al is removed, but the Ge inclusions in the droplets are still visible. (c) Zoomed view of one of the Al/Ge droplets. (c-1), (c-2) and (c-3) display EDX spectra recorded from the rectangular regions labelled 1, 2, and 3 in (c).

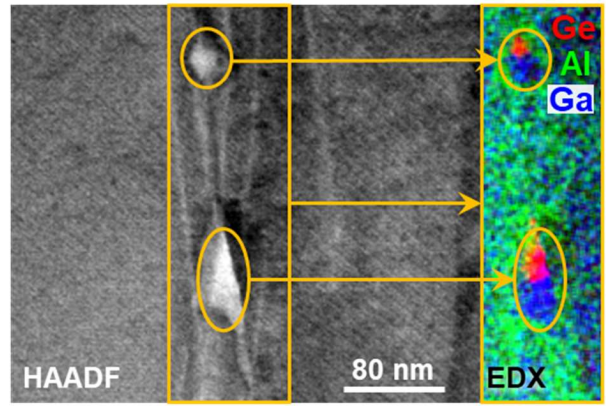

**Figure 3.** In AL40GE2, HAADF-STEM image (left side) and EDX map (right side) of two Ge grains (red) located on top of Ga-rich regions (blue). These crystallites embedded in the AlGaN matrix are much smaller (about 20 nm) than those present at the surface of the sample.

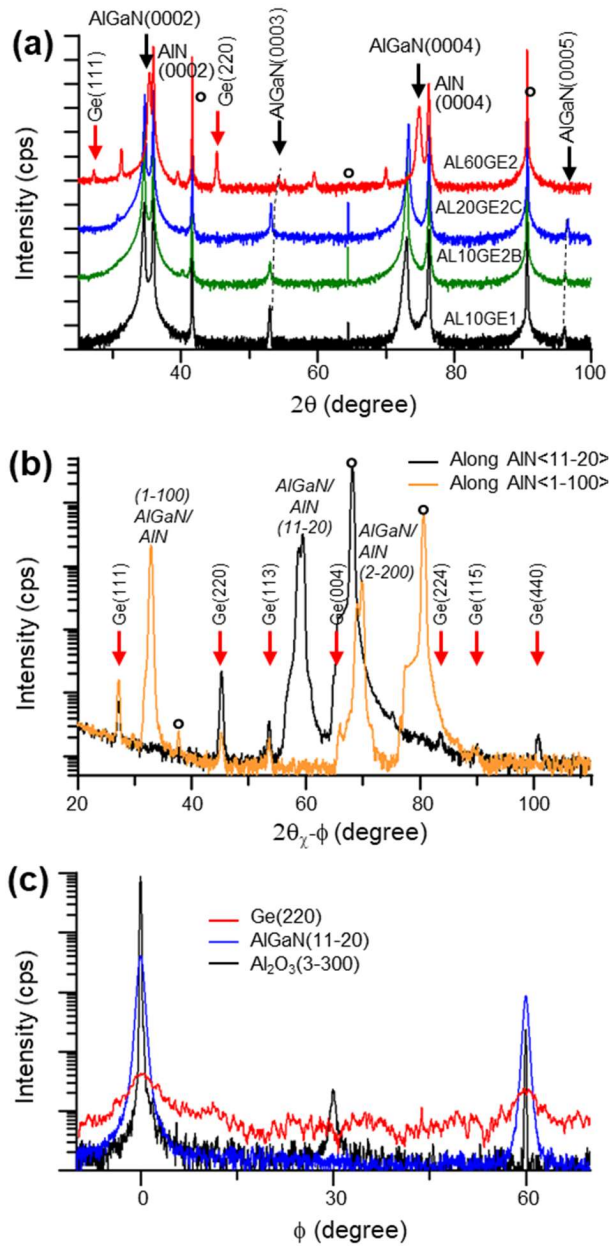

**Figure 4.** X-ray diffraction measurements: (a) Out of plane 2θ–θ scans for 4 of the samples (semilog scale). The curves are vertically shifted for high clarity. $Al_2O_3$ reflections are labeled (°). (b) In-plane scan $2\theta_\chi$–ϕ scan along the <11-20> and <1-100> directions of AlN for sample AL60GE2. (c) ϕ-scan around the Ge(220), AlGaN(11-20) and $Al_2O_3$(3-300) diffraction peaks for sample AL60GE2.

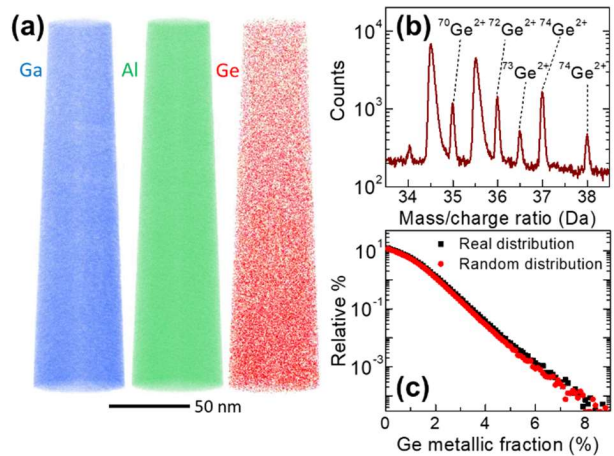

**Figure 5.** From sample AL50GE2: (a) 3D APT reconstruction considering Ga, Al, and Ge ions. The reconstruction was performed assuming a cone angle of 5° and a tip initial radius of 35 nm (data extracted from SEM images of the APT tip before evaporation) (b) Mass spectrum where the peaks corresponding to Ge ions are highlighted. (c) Distribution of Ge composition measured experimentally, compared to the expected distribution in a random alloy.

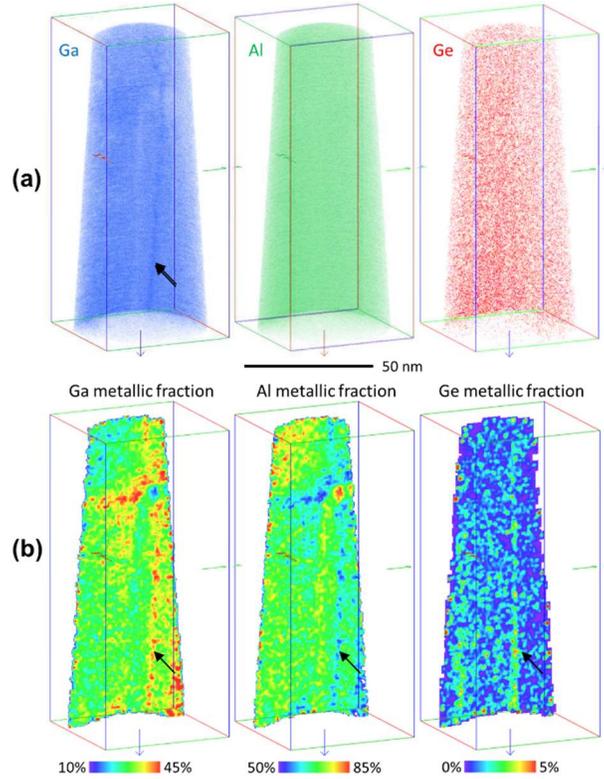

**Figure 6.** From sample AL50GE2: (a) 3D APT reconstruction considering Ga, Al, and Ge ions. The reconstruction was performed assuming a cone angle of 8° and a tip initial radius of 35 nm (data extracted from TEM of the APT tip before evaporation). The arrow in the Ga site reconstruction outlines a vertical Ga-rich region. (b) Vertical composition slice over a 1-nm-thick region. The location of the Ga-rich region correlates with a decrease the Al metallic fraction and an increase of the Ge concentration.

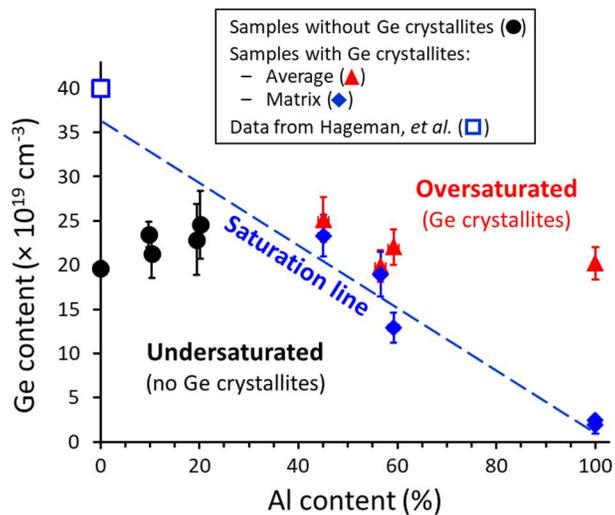

**Figure 7.** Variation of the Ge concentration measured by EDX as a function of the Al mole fraction. Solid symbols are samples in Table 1. The open symbol is extracted from Hageman, *et al.*[26] The dashed line marks the saturation limit of Ge in AlGaN, extracted from a linear fit of the Ge concentration in the matrix in the samples presenting Ge crystallites.

# Solubility limit of Ge Dopants in AlGaN: a Chemical and Microstructural Investigation down to the Nanoscale

C. Bougerol,[†] E. Robin,[‡] E. Di Russo,[§] E. Bellet-Amalric,[‡] V. Grenier,[‡] A. Ajay,[‡,⊥] L. Rigutti,[§] and E. Monroy[‡,*]

[†] Univ. Grenoble-Alpes, Institut Néel-CNRS, 25 av. des Martyrs, 38000 Grenoble, France

[‡] Univ. Grenoble-Alpes, CEA, IRIG, 17 av. des Martyrs, 38000 Grenoble, France

[§] UNIROUEN, CNRS, Groupe de Physique des Matériaux, Normandie Université, 76000 Rouen, France

[*] Corresponding author: eva.monroy@cea.fr

[⊥] Current affiliation: Walter Schottky Institute and Physics Department, Technical University of Munich, 85748 Garching, Germany

## X-RAY DIFFRACTION OF AL100GE2

Figure S1 presents out-of-plane X-ray diffraction 2θ-θ scans of the AL100GE2 sample as grown and after HCL cleaning. For the as grown sample, the diffractogram displays not only the expected AlN reflections, but also very intense peaks originating from pure Al (face-centered cubic phase). After HCl cleaning all the extra peaks disappear.

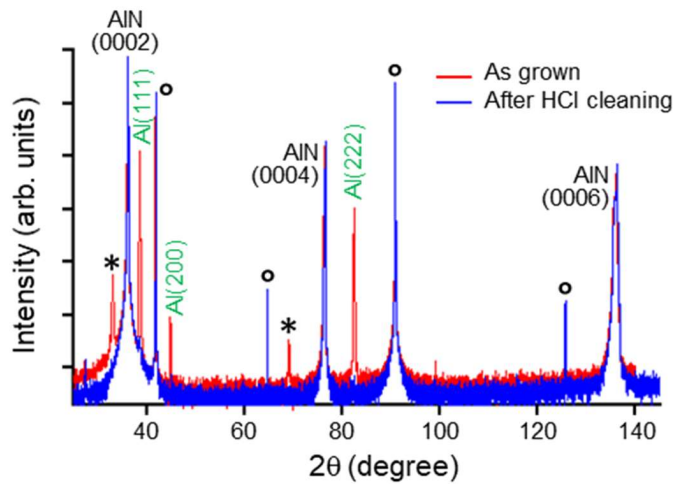

**Figure S1.** X-ray 2θ–θ scan from sample AL100GE2 before and after the post-growth HCl cleaning. The intensity is plotted in semi-log scale, in arbitrary units. The peaks stemming from the sapphire substrate are marked (°). The origin of the 2 peaks labeled (*) is not identified, but they correspond to the same plane family.

## ATOM PROBE TOMOGRAPHY OF AL20GE2C

Figure S2a presents atom probe tomography measurements of a specimen extracted from sample AL20GE2C. The tip contains a threading dislocation that appears as a linear defect with higher Ga mole fraction (between the arrows in the Ga site reconstruction). The Ge distribution presents also a linear Ge-rich trace, which confirms that some threading dislocations are decorated with Ge. Away from the dislocation, the distribution of Ge in the AlGaN matrix is homogeneous, as confirmed by statistical analysis displayed in Figure S2b and compared to an artificial alloy with Ge, Ga and Al atoms mixed in a random manner.

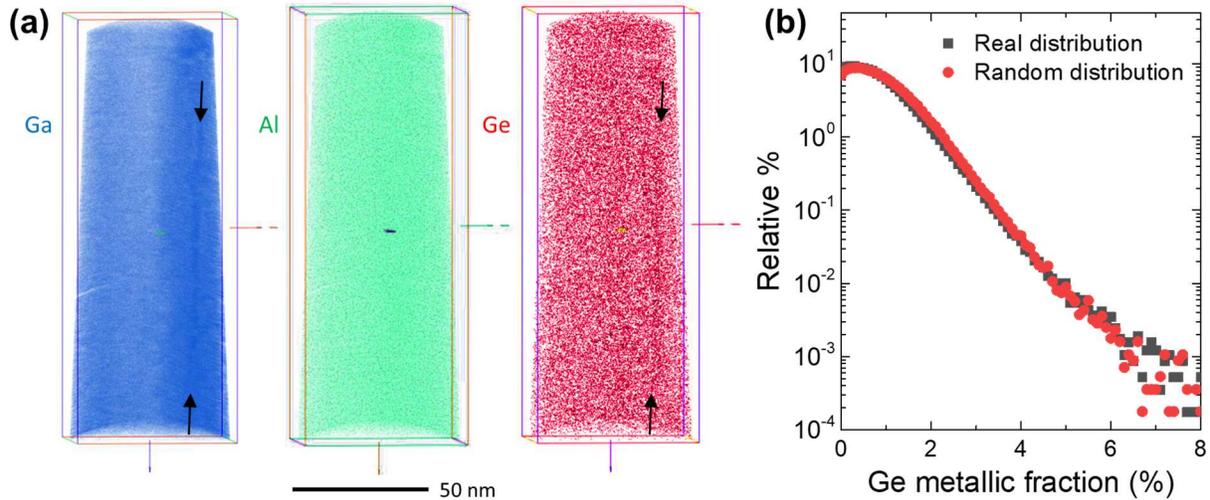

**Figure S2.** From sample AL20GE2C: (a) 3D APT reconstruction considering Ga, Al, and Ge ions. The reconstruction was performed assuming a cone angle of 5° and a tip initial radius of 50 nm (data extracted from SEM of the APT tip before evaporation). The arrows in the Ga site and Ge site reconstructions outline a vertical Ga- and Ge-rich line. (b) Distribution of Ge composition measured experimentally, compared to the expected distribution in a random alloy. The analysis was performed in a volume of 20×20×50 nm$^3$ without dislocations.